\documentstyle[12pt,epsfig,amssymb]{article}
\begin{document}
\setlength{\parskip}{0.45cm}
\setlength{\baselineskip}{0.75cm}
\begin{titlepage}
\begin{flushright}
CERN-TH/97-71 \\
\end{flushright}
\vspace{0.6cm}
\begin{center}
\Large
 { \bf Next-to-Leading Order Analysis of Inclusive and Semi-inclusive Polarized Data}
\vspace{1.2cm}

\large 
 D.\ de Florian$^a$, O.A.\ Sampayo$^b$, R.\ Sassot$^c$
 
\vspace*{1.5cm}
\normalsize
{\it $^a$Theoretical Physics Division, CERN, CH-1211 Geneva 23, Switzerland}

\vspace*{0.1cm}
{\it $^b$Departamento de F\'{\i}sica, 
Universidad Nacional de Mar del Plata \\ 
Funes 3350, (7600) Mar del Plata, Argentina}

\vspace*{0.1cm}
{\it $^c$Departamento de F\'{\i}sica, 
Universidad de Buenos Aires \\ 
Ciudad Universitaria, Pab.1 
(1428) Bs.As., Argentina}

\large
{\bf Abstract}
\end{center}
\vspace{0.5cm}
\normalsize
\noindent
 We present a combined next-to-leading order QCD analysis to
data on both inclusive and semi-inclusive polarized deep inelastic
scattering asymmetries. Performing NLO QCD global fits with different
sets of observables, we evaluate the impact of the very recent semi-inclusive results presented by SMC in the extraction of NLO polarized parton
distributions.

\begin{flushleft}
CERN-TH/97-71 \\
April 1997
\end{flushleft}
  
\end{titlepage}
\newpage

\section{ Introduction}

In recent years, considerable attention has been paid to polarized deep 
inelastic scattering experiments, to the interpretation of the corresponding
data in the framework of perturbative QCD, and to the phenomenological
extraction of non-perturbative spin-dependent parton distributions
\cite{ELLIOT,FORTE,MALLOT}.

The intense activity around these issues have come not only from the
interesting developments and discussions that have arisen in
each of them, but also from the fact that, combined, they are the most
appropriate tools to unveil the
spin structure of nucleons, a subject that is still being debated.

In fact, an increasing amount of high-precision totally inclusive
data, collected by different collaborations over the last few years
[4-18], combined with the recent computation of the complete
perturbative QCD corrections up to next-to-leading order of
the inclusive cross sections \cite{MERTIG,WERNER}, have lead to
several QCD analyses and also extractions of polarized parton
distributions  [15,20-23]. However, many of the results
obtained in those analyses, and particularly in the derivation of
parton distributions, depend strongly on non-trivial
assumptions, which seem to be unavoidable until additional data are
available.

One of the sources foreseen for additional data that can be included in
those analyses is the so-called semi-inclusive spin-dependent
asymmetries. These asymmetries are particularly sensitive to specific
combinations of partons of different flavours and nature, and have been
proposed and used to study the valence-quark distributions in the
proton \cite{SMCD}. Even though this kind of data have been available
for some time \cite{EMCB,SMCD}, it had limited statistics and up to now
only $Q^2$-independent analyses have been performed on it.

More recently, a large amount of more accurate semi-inclusive data have
been produced, and also the appropriate perturbative tools for their
analysis have been developed. The new SMC data \cite{SMCX}, which cover the
same kinematical range as given by the inclusive measurements, superseded previous
presentations with reduced uncertainties. From a more
theoretical point of view, the complete NLO QCD corrections to 
spin-dependent semi-inclusive asymmetries have recently been computed in a
consistent factorization scheme \cite{NPB1,PLB}. There,  NLO effects have
 been estimated, in particular for some observables
originally proposed to disentangle valence-quark contributions, and the
effects of different  kinematical cuts have been
analysed.

In this way, the new data not only allow a more comprehensive analysis
of polarized deep inelastic scattering, but also provide   
a precise test ground for the recently proposed framework for the computation of
higher-order corrections in semi-inclusive processes. Consequently, in
this paper we first evaluate the effect of including the available
semi-inclusive data in global LO and NLO QCD analyses, sum rules
estimates, and parton distribution functions. In this task, we pay special
attention to the release of different constraints usually assumed to be
valid, such as flavour symmetry relations in the estimates of the first moments of the
distributions. Then, we analyse the constraining power of the
semi-inclusive data on the  parton distributions, and
finally we  make definite predictions for the forthcoming
 experiments \cite{HERMES1,COMPASS}.

\section{ Definitions}

In order to fix notation and conventions,  we summarize in this section
the expressions for the LO and NLO inclusive and semi-inclusive spin-dependent asymmetries. These asymmetries are written in terms of
polarized parton distributions, fragmentation and fracture functions,
with the corresponding coefficient functions, defined within a definite
factorization prescription.

For the totally inclusive case, the spin-dependent asymmetries are
given by \cite{ELLIOT}:  
 \begin{equation} A_{1}^{N}(x,Q^2) \simeq
\frac{g_{1}^{N}(x,Q^2)}{F_{1}^N(x,Q^2)}=
\frac{g_{1}^{N}(x,Q^2)}{F_{2}^N(x,Q^2)/\{2x[1+R^N(x,Q^2)]\} } ,
\end{equation}
 where the inclusive spin-dependent nucleon structure
function $g_{1}^{N}(x,Q^2)$ can be decomposed into convolutions between
parton densities $\Delta q_i(x,Q^2)$, $\Delta g(x,Q^2)$, and coefficient functions $\Delta
C_i(x)$:
\begin{eqnarray}
 g_{1}^{N}(x,Q^2)=\frac{1}{2} \sum_{q, \bar q} e_q^2
\left \{ \Delta q (x,Q^2)  
+ \frac{\alpha_s(Q^2)}{2\pi} [ \Delta C_q \otimes
\Delta q  +  \Delta C_g \otimes
\Delta g ] \right\} ,
\end{eqnarray}
where the convolution product is defined by 
\begin{eqnarray}
  \Delta C_f \otimes \Delta f(x,Q^2) \equiv \int_x^1 \frac{dz}{z} \Delta C_f (z)  
\Delta f\left(\frac{x}{z},Q^2 \right) .
\end{eqnarray}
It is customary to define the
coefficient functions in either the usual $\overline{MS}$ scheme or in
other schemes with different factorization properties  \cite{WERNER2}.
In the $\overline{MS}$  scheme, used throughout the present analysis,
the coefficients are given by:  
\begin{eqnarray} \Delta C_q(x)=C_F
\left [ (1+x^2) \left ( \frac{\ln(1-x)}{1-x}\right )_+
-\frac{3}{2}\frac{1}{(1-x)_+}\right.\nonumber \\ \left.
-\frac{1+x^2}{1-x}\ln x +2+x- \left ( \frac{9}{2} +\frac{\pi^2}{3}
\right ) \delta (1-x)\right ] \nonumber \\ \Delta C_g(x)=\frac{1}{2} \left [
(2x-1) \left (  \ln \frac{1-x}{x}-1 \right ) +2(1-x)\right ] .
\end{eqnarray}
 A more detailed discussion about these, including their Mellin moments in 
different factorization  schemes, can be found in
Ref. \cite{GRSV}.

Analogously, for the  semi-inclusive asymmetries, the full NLO
expression can be written as: 
  \begin{equation} \left.
A_{1}^{N\,h}(x,Q^2) \right|_Z \simeq \frac{\int_{Z} dz\,
g_{1}^{N\,h}(x,z,Q^2)}{\int_{Z} dz\,F_{1}^{N\,h}(x,z,Q^2)} ,
\end{equation} 
where the superscript $h$ denotes the hadron detected in
the final state, and  the variable $z$ is given by the ratio between the hadron
energy and that of the spectators in the target ($z=E_h/[E_N\,(1-x)]$, with the
 energies given in the $\gamma^* p$ CM frame).
The region $Z$, over which $z$ is integrated, is determined by
kinematical cuts applied when measuring the asymmetries.  These are
applied in order to suppress target fragmentation contributions and
are often given in terms of lower limit in the variable $z_h=P\cdot h/P\cdot q$.

The semi-inclusive spin-dependent structure function
$g_{1}^{N\,h}(x,z,Q^2)$ can again be decomposed into convolutions
between parton densities $\Delta q_i(x,Q^2)$, $\Delta g(x,Q^2)$, unpolarized fragmentation
functions $D_{h/j}(z,Q^2)$,  coefficient functions $\Delta C_{ij}$, and
polarized fracture functions $\Delta M^h_{i}(x,z,Q^2)$, the latter being given by the 
contribution to the target fragmentation region \cite{NPB1} as
\newpage
\begin{eqnarray} 
g_{1}^{N\,h}(x,z,Q^2) &=& \sum_{q,\overline{q}} e_q^2 \left
\{ \Delta q_i (x,Q^2)  D_{h/i}(z,Q^2) +\frac{\alpha_s(Q^2)}{2\pi}[ \Delta q_i
 \otimes \Delta C_{ij} \otimes D_{h/j} \right.  \nonumber \\
&+& \left. \Delta
q_i  \otimes \Delta C_{ig} \otimes D_{h/g}+ \Delta g  \otimes
\Delta C_{gj} \otimes D_{h/j} ] \right.  \\ 
&+& \left. \Delta M_{q_i}^h (x,z,Q^2)
+\frac{\alpha_s(Q^2)}{2\pi}[\Delta M_{q_i}^h \otimes \Delta C_{i}+\Delta
M_{g}^h \otimes \Delta C_{g}] \right \} \nonumber .
\end{eqnarray} 
A complete computation
of this kind of observable and the full expressions for the
corresponding coefficient functions in different factorization schemes
can be found in Ref. \cite{NPB1}.  An analogous expression can be
written for the unpolarized semi-inclusive structure function
\cite{GRAU}.

In order to be consistent with the factorization prescription chosen for
the inclusive asymmetries in Eq. (3), the following counterterms
for the semi-inclusive expressions have to be used
\begin{eqnarray} 
\Delta \tilde{f}_{q}^{F}(u,\rho)&=& 4(u-1)\, \delta(1-\rho)
\nonumber \\  
\Delta \tilde{f}_{q}^{MI}(u,\rho) &=&
4(u-1)\, \delta(\rho-a)\nonumber \\ 
\Delta \tilde{f}_{q}^{MH}(u) &=& 4(u-1)
\nonumber \\ 
\Delta \tilde{f}_{g}^{F}(u,\rho)&=& 0 \nonumber \\ \Delta
\tilde{f}_{g}^{MI}(u,\rho)&=&0 \nonumber \\ \Delta
\tilde{f}_{g}^{MH}(u,\rho)&=&0 
\end{eqnarray}
in the expressions of Ref. \cite{NPB1}.

\section{Hadronization and Input Distributions}

The expressions for the semi-inclusive asymmetries given in the last
section clearly show that the analysis of these asymmetries requires
not only some knowledge of the unpolarized structure function
$F^{N}_{1}(x,Q^2)$, as in the totally inclusive case, but also of
details about the hadronization processes. These details come  mainly
through the unpolarized fragmentation functions $D_{h/i}(z,Q^2)$, which
are present in both semi-inclusive structure functions
$g^{N\,h}_{1}(x,Q^2)$ and $F^{N\,h}_{1}(x,Q^2)$, and also from fracture
functions \cite{PLB}.

Charged pion and kaon fragmentation functions have been measured in
different experiments, and the corresponding LO and NLO parametrizations
have also been obtained \cite{FRAG1,FRAG2}. In our computations we use
those of Ref. \cite{FRAG2}  and a parametrization of semi-inclusive EMC data
\cite{EMCpi} in order to distinguish between  favoured and unfavoured distributions.
The assumption of  SU(3) symmetry for the sea 
distributions introduces  negligible corrections for the charged 
asymmetries, but very large ones for the difference asymmetries.
 Although the main contributions to charged-particle fragmentation
 come from pions, we also include those related to kaons for
 completeness.

Unpolarized parton densities enter the analysis directly in the
normalization of the inclusive asymmetries, and also convoluted with
fragmentation functions in the semi-inclusive ones. 
At variance with the inclusive case, where the unpolarized observables
$F_2$ and $R$ used to obtain $F_1$ can be taken directly from the data,
in the semi-inclusive case, these have to be computed using the parton distributions. Consequently, and in order to be consistent, throughout the present analysis all the unpolarized observables are constructed  using the parton distributions of Ref. \cite{GRV} in their LO and NLO
($\overline{MS}$) versions, according to the order of the fit, and with the 
appropriate QCD coefficients. In particular, this means that $R$ is equal to zero at LO and is given by the corresponding perturbative expression at NLO.
We also use GRV parton distributions in order to check the positivity
constraints on polarized distributions, and the $\Lambda_{QCD}$ values
obtained in that analysis.

Polarized and unpolarized fracture functions \cite{VEN,GRAU,NPB1}
describe the details of hadro\-ni\-za\-tion processes coming mainly
from target fragmentation region. Although their inclusion is
crucial in order to consistently factorize collinear divergences, once
this process is through, their actual contribution to the cross sections
can be be suppressed by imposing the appropriate kinematical cuts \cite{PLB}.
  Consequently, we
restrict our analysis to single asymmetries for $z_h>0.2$,
leaving for the moment the discussion of difference asymmetries, and
neglecting fracture function contributions. Eventually, high-precision
semi-inclusive experiments will allow accurate extractions of these
distributions.

\section{ Initial Parton Distributions}

Over the last couple of years, several NLO QCD global fits to data on
totally inclusive polarized asymmetries have been presented
[15,20-23]. The approaches implemented in each of these
analyses generally differ not only according to the set of data available when
they were performed, but also to the functional dependence, initial
scale, and factorization prescription chosen for the input parton
distributions, in analogy to what happens in spin-independent
analyses.

However, at variance with what is found in the latter case,
spin-dependent data allow equally good fits, i.e. with similar values
of $ \chi ^2 /$d.o.f., but with parton distributions rather different in
shape and normalization, even within the measured region. These
differences are moderated for valence-quarks distributions, but rather
large for sea quarks and gluons. A suggestive example of this, is given by the
differences between the gluon normalizations of the most recent
analyses \cite{ABFR,E-154}, even though both have been performed in the same
 AB factorization scheme and with almost the same data.   
In general,
the fitting procedure prefers one set or another depending very strongly on
the functional form of the initial parton distributions, and some additional constraints
imposed over the distributions, such as positivity, flavour symmetry,
or even more arbitrary assumptions, which may be freely chosen (with
no significant consequence in the value of $\chi^2/$d.o.f.). 

Consequently, although most of the analyses show some common global
features, such as a non-negative and not very large polarized gluon
density, regarding the extraction of polarized parton distributions, we
are far  from the accuracy attained in the unpolarized case; then, 
more inclusive data and new measurements will be necessary.
In the mean time, in order to design useful experiments and make
predictions for these new observables, we need parton distributions
covering the wide range of possibilities allowed by present data.

These are the main reasons for which, in the present analysis, rather
than adopting some or other stringent constraint on the normalization
of the valence, sea quarks, or gluon densities,  then singling out
the set that presents the lowest $\chi^2$ (given those and other less
apparent assumptions), we adopt a more flexible scheme for the valence
and sea sectors, we put greater emphasis on the measured region, and we
explore different gluon possibilities. It should be noticed that the
usual constraints over the normalizations can in turn  introduce a significant
dependence on the functional behaviour assumed for the
unmeasured region, and fix the values for the sum-rule estimates.

At variance with other parametrizations, we also include in our study
the NLO analysis of semi-inclusive data, which is in principle specially sensitive
to the valence sector and allows a further constraint on them. It is worth stressing  that in this case it is not enough to deal with only quark-singlet and nonsinglet distributions as in the inclusive case \cite{ABFR}. In order to construct  the semi-inclusive observables each flavour distribution has to be individualized.
As we are primarily interested in the measured region, we adopt a rather
simple parametric form for the input spin-dependent valence quark
densities: 
 \begin{equation}
 x \Delta q_V (x,Q_0^2)=N_{q_V}
\frac{x^{\alpha_q}(1-x)^{\beta_q}(1+\gamma_q\,
x)}{B(\alpha_q+1,\beta_q+1)+\gamma_q \, B(\alpha_q+2,\beta_q+1) },
\end{equation}
 where the parameters $\alpha_q$ and $\gamma_q$ are
obtained from the fitting procedure, and $\beta_q$ is externally fixed
by the positivity constraint with respect to GRV unpolarized parton
distributions at large $x$. 
($\beta_u=3.00(3.33)$ and $\beta_d=3.95(4.26)$ at LO(NLO)).
The initial scale $Q^2_0$ is chosen to be
$0.5\, $GeV$^2$, which is sufficiently low as to induce through the
evolution a more complex and appropriate $x$-dependence at higher
scales. We have also tried  different choices for the initial scale, finding very similar results for quarks but significant changes in the gluon density. This reflects a large uncertainty on the gluon distribution, not only regarding the $x$-dependence, but also on its first moment.

In order to trace and parametrize the departure from the SU(2)
and SU(3) flavour symmetries, we define the normalization
coefficients $N_{q_V}$ in terms of the $F$ and $D$ constants and two
additional  parameters. In this respect,
it is customary to relate the first moment of the input parton
densities to the $F$ and $D$ constants through relations like 
\footnote{ The $\delta$ notation means that  the first moment of 
the polarized  distribution has been taken.}  
\begin{equation} 
\delta u +\delta \overline{u}- \delta d -\delta
\overline{d} = F+D 
\end{equation} 
\begin{equation} 
\delta u +\delta
\overline{u}+ \delta d +\delta \overline{d}- 2(\delta s +\delta
\overline{s}) = 3F-D .
\end{equation}
 Imposing additional symmetry
relations such as $\delta \overline{u}=\delta \overline{d}$  Eq. (9) becomes
 \begin{equation}
\delta u_V - \delta d_V = F+D
 \end{equation}
 and making $\delta
\overline{u}=\delta \overline{d}=\delta \overline{s}$ Eq. (10)
turns into
 \begin{equation}
 \delta u_V + \delta d_V = 3F-D .
\end{equation} 
Equations (11) and (12) completely fix the valence quark
normalizations.  These relations, although they are sensible approximations, may
not be true, and their enforcement  strongly depends on the
unmeasured  low-$x$ behaviour of the densities.  In order to relax
these restrictions we propose:  
\begin{equation} 
\delta u_V - \delta
d_V = (F+D)(1+\epsilon_{Bj}) 
\end{equation} 
and 
\begin{equation}
 \delta u_V + \delta d_V  + 4 (\delta \overline{u}-\delta s ) =
(3F-D)(1+\epsilon_{SU(3)}).
 \end{equation}
 The parameters
$\epsilon_{Bj}$ and $\epsilon_{SU(3)}$ account quantitatively for
eventual departures from  flavour symmetry considerations (including
also some uncertainties on the low-$x$ behaviour). They also measure the
degree of fulfilment of the Bjorken  \cite{BJ} and Ellis-Jaffe sum rules
\cite{EJ}.

For the light quarks (for simplicity $\Delta \overline{u}=\Delta \overline{d}$ is assumed throughout this paper) the proposed input density is given by:
\begin{equation}
 x\Delta \overline{q} (x,Q_0^2)=N_{\overline{q}}
\frac{x^{\alpha_{\overline{q}}}(1-x)^{\beta_{\overline{q}}}}
{B(\alpha_{\overline{q}}+1,\beta_{\overline{q}}+1)},
\end{equation}
 where  $\alpha_{\overline{q}}$, $\beta_{\overline{q}}$,
and $N_{\overline{q}}$ are  only constrained by
positivity. The same functional dependence and considerations are used
for gluons, since using more pa\-ra\-me\-ters seems to be useless, taking into 
account the uncertainties on them. For strange quarks we adopt: 
 \begin{equation}
 \Delta \overline{s} (x,Q_0^2)=N_{\overline{s}}\, \Delta \overline{q} (x,Q_0^2),
\end{equation} 
finding pointless the addition of more parameters.


\section{ Results}

In the following we report  the results obtained from several global
fits performed with different sets of data and also varying the
constraints imposed over the parton densities and the order of
perturbation.

 Throughout the present analysis, we consider as totally inclusive  data for proton targets
 the results presented in refs. \cite{EMCB,E-143A,E-143C,SMCNEW}, for deuteron targets 
those in \cite{SMCE,E-143A,E-143C}, and for neutron targets those in  \cite{E-142,HERMES,E-154}.
 In order to avoid possible higher-twist contributions, we have
taken into account only measurements with $Q^2>1\,$GeV$^2$ given a total of 133 data points. As
semi-inclusive data we take those recently presented by SMC \cite{SMCX}, 48
data points, which then lead to combined global fits with 181 data points.
Correlations between totally-inclusive and semi-inclusive SMC data sets
have been taken into account, and increase the total $\chi^2$.

\begin{center}
\begin{tabular}{|c|c|c|c|c|c|c|} \hline \hline
 {\footnotesize Parameter}    &  \multicolumn{3}{c|}{NLO {\footnotesize ($\overline{MS}$)} }&
   \multicolumn{3}{c|}{LO} \\ \cline{2-7}
    & {\footnotesize Set 1}  & {\footnotesize Set 2} &{\footnotesize
    Set 3}  & {\footnotesize  Set 1 } & {\footnotesize  Set
2} & {\footnotesize Set 3 }\\ \hline\hline $
 \chi^2_{T}$&153.95 & 152.69  & 152.87 & 158.77 & 157.64 & 159.92\\ \hline 
$\chi^2_{I}$&101.90 & 100.47        & 100.84 & 107.56 &  106.37      & 108.73 \\ \hline
$\chi^2_{SI}$&44.62 &  45.64     &  45.24 & 44.70  & 44.56    & 44.13  \\ \hline\hline
$\epsilon_{Bj}$ & $-$0.019 & $-$0.021 & $-$0.023    & $-$0.037 & $-$0.045 & $-$0.035 \\ \hline 
$\epsilon_{SU(3)}$ & $-$0.10 & $-$0.10 & $-$ 0.10    & $-$0.10 & $-$0.10 & $-$0.098 \\ \hline
$\alpha_u $ & 0.896& 0.888 & 0.895   & 0.762 & 0.787 & 0.75\\ \hline 
$\gamma_u $ & 6.68 & 6.92 &  6.73 & 7.71 & 7.04 & 8.17\\ \hline
$\alpha_d $ & 0.69 & 0.71 & 0.688  & 0.61 & 0.62 & 0.56\\ \hline 
$\gamma_d $ & 11.18 & 11.53 &  12.22 & 6.24 & 7.67 & 9.73\\ \hline 
$N_{\overline{q}}$ & $-$0.054 & $-$0.051 & $-$0.045    & $-$0.053& $-$0.049 & $-$0.043\\ \hline
$\alpha_{\overline{q}} $ & 0.70 & 0.70 & 0.70 & 1.0 & 1.0 & 1.0\\ \hline
$N_{g} $ & 0.80 & 0.40 &  0.10  & 0.85 & 0.48 & 0.10 \\ \hline
$\alpha_{g} $ & 1.08 & 2.80 & 2.00   & 1.41 & 2.29 &  2.00
\\ \hline $\beta_{g} $ & 6.00 & 9.10 & 6.00    & 10.59 & 13.52  & 12.71 \\ \hline \hline
\end{tabular} 

\vspace*{5mm} {\bf Table 1: Combined
global fits}. \end{center}

In Table 1 we show the results for three different NLO ($\overline{MS}$)
 and LO global fits for combined inclusive and
semi-inclusive data in which the gluon density first moments $N_g$ are
constrained  to three different regions:
\begin{eqnarray}
\rm{Set\, 1} &&\,\,\,\,\,\,\,\,\,\,\,\,\,\,\,\,\,\delta g > 0.8\, \nonumber \\
\rm{Set\, 2} && \,\,0.1 > \delta g  > 0.8 \nonumber \\
\rm{Set\, 3} &&\,\,\,\,\,\,\,\,\,\,\,\,\,\,\,\,\,\delta g < 0.1 , \nonumber 
\end{eqnarray}
 defined at the initial scale. The breaking parameter
 $\epsilon_{Bj}$  is left free whereas,   $\epsilon_{SU(3)}$  is constrained to allow only
 moderate violations of the polarized sum rules. Since this last parameter is not well determined by the data, we allow it to vary between $-$0.1 and 0.1 as a compromise between data and theoretical expectations; when left free it varies between $-5\%$ and $-40\%$ without modifying significantly the $\chi^2$ value. Therefore it is  not possible yet to determine accurately the nonsinglet axial current $a_8$ from the existing data.

The table does not include the values for the $\beta_{\overline{q}}$ and $N_s$ parameters; the first one was found to be constrained by positivity to 7.80 and 
6.10, at NLO and LO respectively. Regarding $N_s$, although the strange-sea normalization is allowed to vary with respect to the one of the light quarks, the fits favour almost the same value, so we fix it to be equal to 1. 

The first row in Table 1 shows the best $\chi^2$ values obtained in each of the
three allowed regions for the gluon normalization, both in NLO and in LO, taking
into account both sets of data  (181 data points). The  following two rows discriminate
between the contributions to the total $\chi^2$ coming from the inclusive and 
semi-inclusive data sets, respectively (133 and 48 points). Clearly, the semi-inclusive data set is in  very good agreement  with the inclusive one, and allows fits of remarkable quality in the three gluon regions.

In the combined fits there is a   preference for sets with a   moderate gluon polarization, which is reflected in the saturation of the constraints imposed on the gluon normalization in the case of sets 1 and 3.
However, the differences in $\chi^2$ values obtained in each of the regions are so subtle that the uncertainty in the value for the first moment of the polarized gluon density is significantly large, and even a slightly negatively polarized distribution for gluons can not be ruled out yet.

In Fig. 1 we compare the inclusive asymmetries coming from Set 2 (NLO and LO, respectively) with the data. The lines interpolate the fit estimates at the
mean $x$ and $Q^2$ values quoted by the different experimental collaborations. 
As can be seen, the differences between NLO and LO fits are significant only in the region of large $x$,  where data have larger error bars. The  estimates coming from the remaining sets of parton distributions are not shown, as they lead to almost identical asymmetries. It is apparent from  Fig. 1  that the neutron asymmetry is dominated by the new E-154 data, whereas a combination between E143 and SMC fixes the proton behaviour. 

In Fig. 2 we show the same but for the semi-inclusive data. Notice that the large error bars of these data reduce its weight in the global fit and that the main difference in the $\chi^2$ between LO and NLO fits comes from the totally inclusive data. Also in Fig. 2 we show the result of a fit using only the semi-inclusive data as described below. 

In Tables 2 and 3 we show sum rules and first moments estimates  for the three sets at different scales. For the Bjorken sum rule $\Gamma^{Bj}$, the departure from the theoretical expectation is significantly small, as given by the small values found for the parameter $\epsilon_{Bj}$.

\begin{center}
\begin{tabular}{|c|c|c|c|c|c|c|c|c|c|} \hline \hline
{\footnotesize Fit} &  {\footnotesize $Q^2$}  & $\Gamma_1^p$ & $\Gamma_1^n$ & $\Gamma^{Bj}$ & $\delta\Sigma$ & $\delta g$ & $\delta u_V$ & $\delta d_V$ & $\delta \overline{q}$ \\ \hline \hline
Set 1 & 1 & 0.123 & $-$0.059& 0.183 & 0.194 & 1.12& 0.876 & $-$0.356 & $-$0.054 \\ \cline{2-10}
& 4      & 0.127 & $-$0.062& 0.189 & 0.190  & 1.69& 0.875 & $-$0.355 & $-$0.054 \\ \cline{2-10}
     & 10 & 0.129 & $-$0.063& 0.192 & 0.190 & 2.02& 0.874 & $-$0.355 & $-$0.054 \\ \hline 
Set 2 & 1 & 0.124 & $-$0.057& 0.182 & 0.212 & 0.59& 0.875 & $-$0.354 & $-$0.051\\ \cline{2-10}
      & 4 & 0.129 & $-$0.060& 0.189 & 0.207 & 0.91& 0.874 & $-$0.354 & $-$0.052\\ \cline{2-10}
     & 10 & 0.130 & $-$0.061& 0.191 & 0.206 & 1.11& 0.873 & $-$0.354 & $-$0.052\\ \hline 
Set 3 & 1 & 0.128 & $-$0.054& 0.182 & 0.247 & 0.19 & 0.874 & $-$0.353 & $-$0.046 \\ \cline{2-10}      
      & 4 & 0.132 & $-$0.056& 0.189 & 0.242 & 0.34 & 0.873 & $-$0.352 & $-$0.046 \\ \cline{2-10}
     & 10 & 0.135 & $-$0.057& 0.191 & 0.240 & 0.43 & 0.872 & $-$0.352 & $-$0.046 \\ \hline \hline
\end{tabular} 

\vspace*{5mm} {\bf Table 2: Sum rules from NLO combined fits}. \end{center}

As usual in the $\overline{MS}$ scheme,
the first moment of the singlet distribution, $\delta \Sigma$, is found to be considerably smaller than the naive prediction, and is correlated to the gluon polarization. Notice that the valence-quark normalizations are quite stable and give the same result, independently of the singlet sector and that in the case of the polarized sea we show the first moment corresponding to $u$ and $d$ quarks, being negligible the differences with the one of $s$ quarks.

\begin{center}
\begin{tabular}{|c|c|c|c|c|c|c|c|c|c|} \hline \hline
{\footnotesize Fit} &  {\footnotesize $Q^2$}  & $\Gamma_1^p$ & $\Gamma_1^n$ & $\Gamma^{Bj}$ & $\delta\Sigma$ & $\delta g$ & $\delta u_V$ & $\delta d_V$ & $\delta \overline{q}$ \\ \hline \hline
Set 1 & 10 & 0.138 & $-$0.064& 0.202 & 0.202 & 2.13& 0.866 & $-$0.344 & $-$0.053 \\ \hline 
Set 2 & 10 & 0.140 & $-$0.060& 0.200 & 0.227 & 1.27& 0.861 & $-$0.340 & $-$0.049 \\ \hline 
Set 3 & 10 & 0.145 & $-$0.057& 0.202 & 0.264 & 0.39& 0.867 & $-$0.346 & $-$0.043\\ \hline \hline
\end{tabular} 

\vspace*{5mm} {\bf Table 3: Sum rules from LO combined fits}. \end{center}

\begin{center}
\begin{tabular}{|l|r|r|r|r|r|r|} \hline \hline
 {\footnotesize }    &  \multicolumn{3}{c|}{NLO {\footnotesize ($\overline{MS}$)} }&
   \multicolumn{3}{c|}{LO}  \\ \cline{2-7}
    & \multicolumn{1}{c|}{\footnotesize Set 1}  &  \multicolumn{1}{c|}{\footnotesize Set 2} & \multicolumn{1}{c|}{\footnotesize
    Set 3}  & \multicolumn{1}{c|}{\footnotesize  Set 1 } & \multicolumn{1}{c|}{\footnotesize  Set
2} & \multicolumn{1}{c|}{\footnotesize Set 3 } \\ \hline\hline 
 $\Gamma_1^p${\scriptsize $(0-0.003)$}&  {  $-$0.006} & { $-$0.002} & {0.001}& {  $-$0.004}& { $-$0.0005}& { 0.003} \\ \hline 
 $\Gamma_1^n${\scriptsize $(0-0.014)$}&  { $-$0.027} & { $-$0.023} & {  $-$0.019}&{  
 $-$0.026}& {  $-$0.020}& {  $-$0.017} \\ \hline
 $\Gamma^{Bj}${\scriptsize $(0-0.014)$} & {   0.026} &{  0.025} & {  0.026}& {  0.027}&{  0.026} &{  0.027} \\ \hline\hline
 \end{tabular} 
\vspace*{5mm} \end{center} \begin{center} {\bf Table 4:   Sum rule extrapolations through the unmeasured region computed at $Q^2=10\, $GeV$^2$.} \end{center}

The first moments of the polarized structure functions, $\Gamma_1^p$ and $\Gamma_1^n$, are in  agreement  with the values estimated by the experimental collaborations even though the asymptotic behaviour of our distributions ($g_1$ goes to very large negatives values at small $x$) is quite different from the Regge expectation assumed in most of the analyses ($g_1 \approx$ constant).   Of course, this behaviour is fixed by the available data at larger $x$ and therefore depends ultimately on the shape assumed for the input parton distributions \cite{ABFR}. This extrapolation
is still the  largest  source of  error for the experimental determination of the sum rules \cite{SMCNEW}. 
As an example, we show  in Table 4, the contributions of the different sets to
the unmeasured regions of the SMC and E154 proton and neutron experiments, respectively.  Notice the large differences between each extrapolated contribution.  In the case of proton target, the extrapolations may even show   opposite signs for different sets and large differences when switching from NLO to LO, due to the fact that NLO gluons  -convoluted with a negative coefficient- contribute directly to the structure function and to differences in the value of $F_1$ used at each order to reconstruct $g_1$ from the asymmetries.

The impact of the semi-inclusive data in the total fit has been estimated  performing also fits using only inclusive data. In these fits we have found that the quark parameters  change less than $2\%$, whereas the changes are a somewhat larger for the gluon distribution. However, the uncertainties already pointed
out about the gluon density dominate over any potential influence of the semi-inclusive data set. The reasons for this very small impact are, basically, the fact that semi-inclusive data has not reached yet the precision and statistical significance of the inclusive one, and also  that the data sets are not completely independent. This can be seen either in the correlations between inclusive and semi-inclusive asymmetries \cite{SMCX}, and also in the fact that parametrizations obtained using only inclusive data give a very good description of the semi-inclusive asymmetries.    

Additionally, it is possible to use the semi-inclusive data in QCD global fits 
but without employing the inclusive data sets directly, for the comparison of 
the corresponding results.  As in this case, not all the parameters can be 
unambiguously fixed by the semi-inclusive data alone,  we have fixed the ones 
corresponding to the gluon and sea densities to the values obtained in Set 2, 
and then adjusted only the valence-quark distributions, with the results shown 
in Table 5.

\begin{center}

 \begin{tabular}{|c|r|r|} \hline \hline
{\footnotesize Parameter}& \multicolumn{1}{c|}{NLO {\footnotesize ($\overline{MS}$)} } &\multicolumn{1}{c|}{LO} \\ \hline \hline
$\chi^2_{SI}$&  40.25 & 39.45\\ \hline 
$\epsilon_{Bj}$ & $-$0.129 & $-$0.131  \\ \hline 
$\epsilon_{SU(3)}$ & 0.088 & 0.076\\ \hline 
$\alpha_u $  & 0.386 & 0.376 \\ \hline 
$\gamma_u $ & 31.69 & 22.81 \\ \hline
$\alpha_d $ & 0.638& 0.565 \\ \hline 
$\gamma_d $ & $-$1.075  & $-$3.363\\ \hline 
$\delta u_V ^*$ & 0.86 & 0.86 \\ \hline 
$\delta d_V ^* $ & $-$0.23  & $-$0.23 \\ \hline \hline
\end{tabular}
\vspace*{5mm} \end{center} \begin{center} {\bf Table 5: Semi-Inclusive
Valence Fits} \\{($^*$ Moments taken at $Q^2=10\, $GeV$^2$).} \end{center}

In these fits,
the $\chi^2$ values with respect to the semi-inclusive data, $\chi^2_{SI}$, are reduced in some units; however, the total $\chi^2$ computed with the obtained distributions increases dramatically to unacceptable values ($\chi^2_{T}>290$), with the largest contributions to it coming from the E-154 neutron data, mainly due to differences in the $\Delta d_V$ distributions obtained from total and SI fits, as  can be seen in Fig. 3, where the parton densities given by the different fits are shown at the common value of  $Q^2=10$ GeV$^2$. 

In the semi-inclusive case, the $\Delta d_V$ distribution is mainly
   constrained by the  deuteron  asymmetry, at variance from the inclusive case, where  is determined by the more accurate E-154 neutron data. As  can be seen in Fig. 2, the difference between  the result for the deuteron asymmetry coming either from the combined fit or  the semi-inclusive one is   apparent, 
even though the  $\Delta d_V$ distributions are quite different, showing the low sensitivity of deuteron observables to this density.  
These obtained $\Delta d_V$'s are of course in agreement when the large errors     coming from the  data (specially the SI set) are taken into account in the corresponding distributions and the same occurs with the first moment, whose central values is found to be smaller than the one obtained in the total analysis mainly due to the change of sign of the SI-distribution at large $x$.

Ongoing semi-inclusive measurements using $^3$He targets can be quite useful
in the determination of valence-quark distributions from semi-inclusive data alone, and also as further constraints in global fits. In Fig. 4 we show predictions for semi-inclusive
production of charged hadrons and $\pi^0$ for $^3$He targets using the combined fit, the one obtained with only semi-inclusive data, and also the prediction coming from the GRSV \cite{GRSV} polarized parton distributions. These asymmetries are particularly sensitive to $\Delta d_V$, which is the main reason for the large differences between the predictions of different sets, specially the one for the production of positively charged hadrons, as can be expected from very simple arguments based on the values of the corresponding fragmentation functions. The 
lines interpolate the $x$ and $Q^2$ values quoted in  the HERMES totally inclusive measurements, and the same cut $z_h>0.2$ has been imposed in order to suppress both target fragmentation effects and final-state mass corrections
(proportional to $4 M_h^2/z^2/W^2)$, which can be significant for low centre-of-mass energy experiments.

\section{ Conclusions}

 Performing a LO and NLO global analysis to both inclusive and semi-inclusive polarized deep inelastic data, we have found that the present semi-inclusive data can be consistently included in global analyses. These global fits show features similar to those coming from totally inclusive data, i.e. a poorly constrained gluon distribution and better determined valence densities, with the semi-inclusive data  introducing  very  small modifications in the  valence 
densities.

The presented LO and NLO polarized parton distributions explore different gluon scenarios  and are therefore very well suited to study  the sensitivity of different observables to the polarized gluon distribution \footnote{A Fortran code with the corresponding parton distributions can be requested from the authors by e-mail at daniel.de.florian@cern.ch}.
 
 Present semi-inclusive data alone fail to define a $\Delta d_V$ distribution 
consistent with those extracted from inclusive data; consequently,  the 
corresponding sets cannot reproduce the inclusive asymmetries for neutron 
targets.  However, ongoing semi-inclusive experiments using  $^3$He targets 
\cite{HERMES}, or more accurate measurements on proton and deuteron targets 
\cite{COMPASS}, can reverse this situation and provide an enhanced perspective 
of the spin structure of the nucleon.

\noindent{ \bf Acknowledgements}

We warmly thank C. A. Garc\'\i a Canal, J. Pretz, M. Stratmann and W. Vogelsang   for interesting discussions.
The work of one of us (D.de F.) was partially supported by the World Laboratory.

\pagebreak


\textheight     9in
\topmargin    -.5in
\textwidth    6.2in
\oddsidemargin 0.4cm
\begin{figure}[htb]
\begin{center}
\mbox{\kern-1cm
\epsfig{file=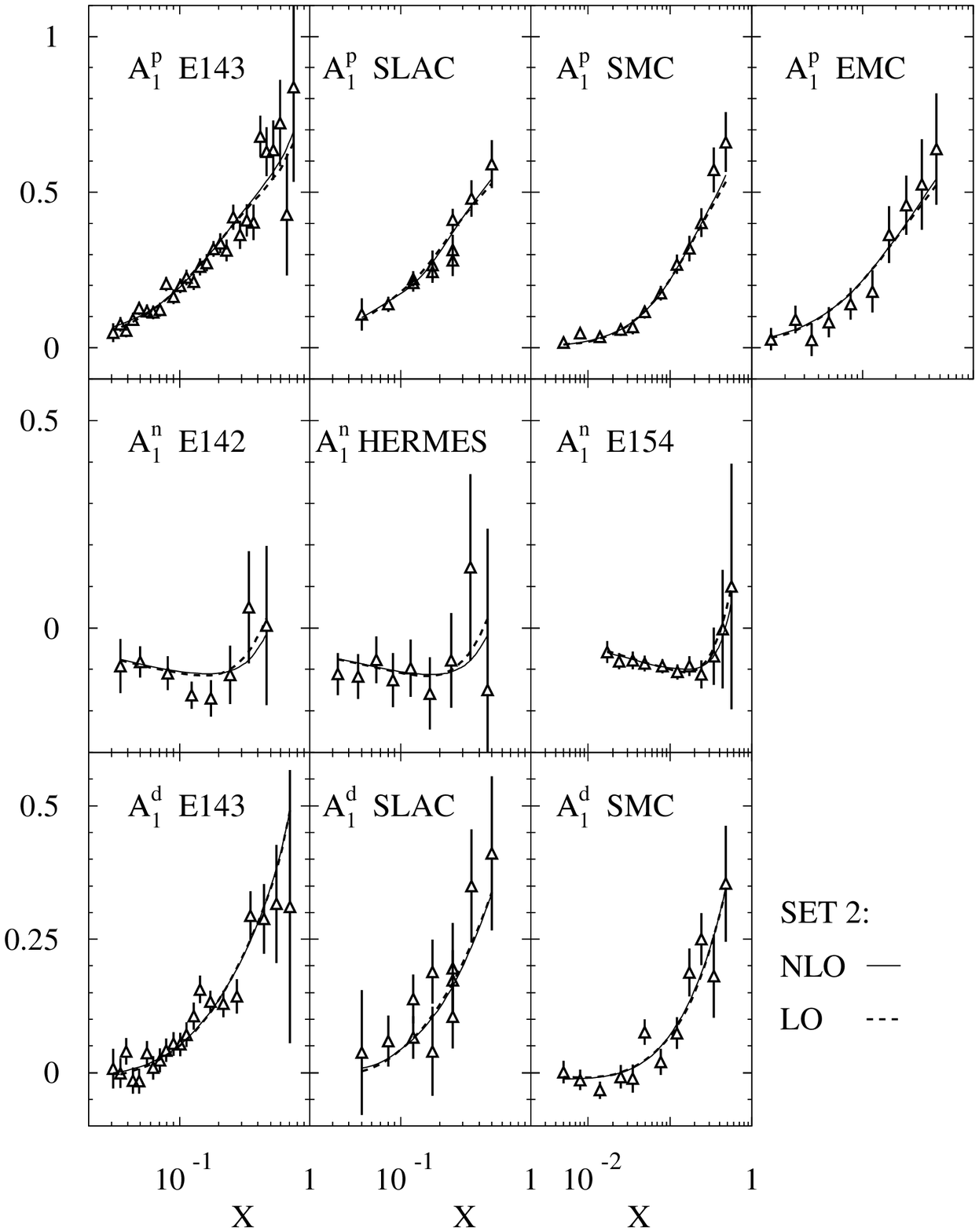,width=17.0truecm,angle=0}}
\caption{ Inclusive asymmetry data against the expectations  from  Set 2 at NLO (solid lines) and at LO (dashed lines).}
\end{center}
\end{figure}
\vskip 14pt
\begin{figure}[htb]
\begin{center}
\mbox{\kern-1cm
\epsfig{file=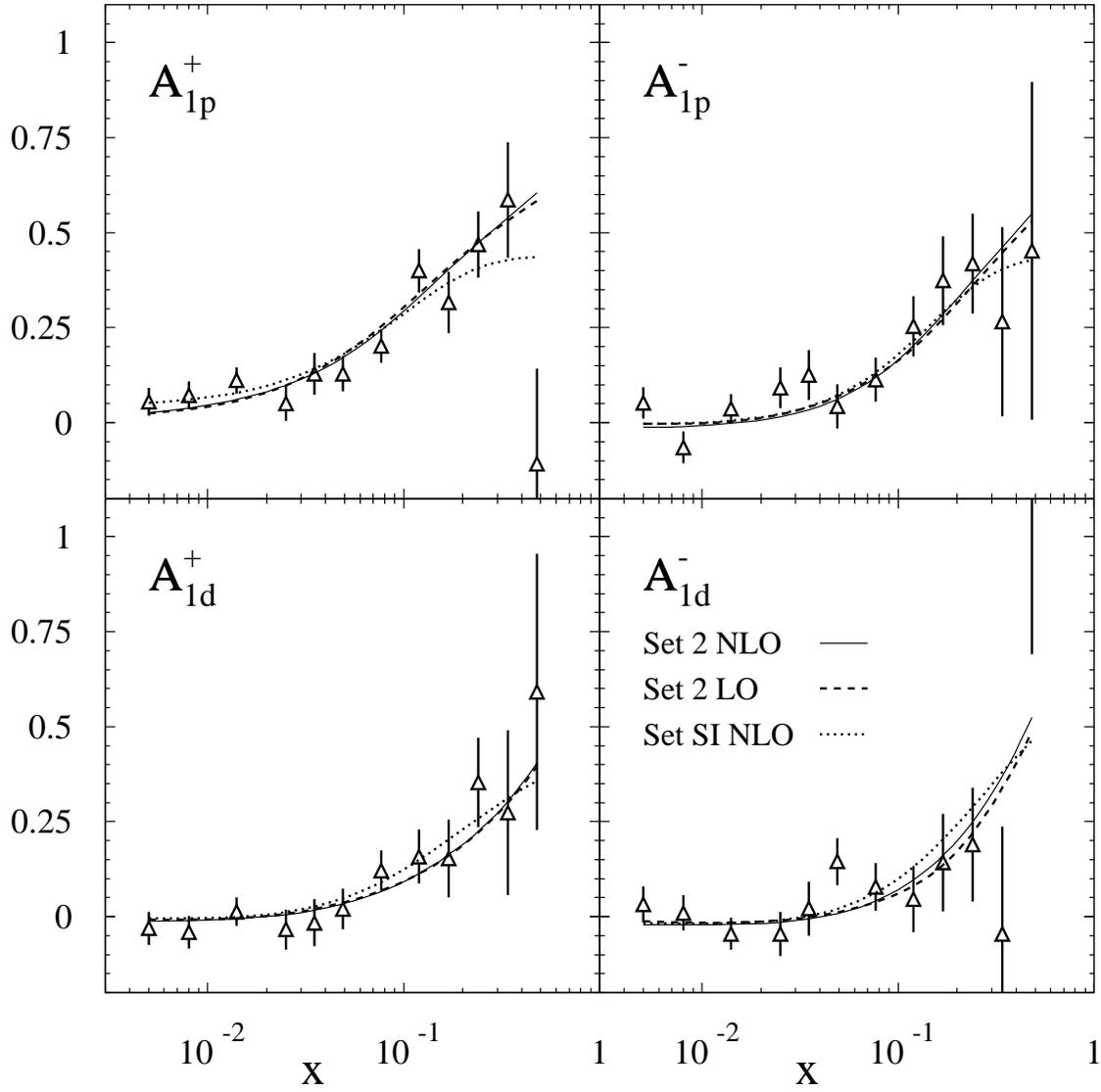,width=17.0truecm,angle=0}}
\caption{
The same as Fig. 1, but for semi-inclusive asymmetries,
and the expectation  from the semi-inclusive set (dots).}
\end{center}
\end{figure}
\vskip 14pt
\begin{figure}[htb]
\begin{center}
\mbox{\kern-1cm
\epsfig{file=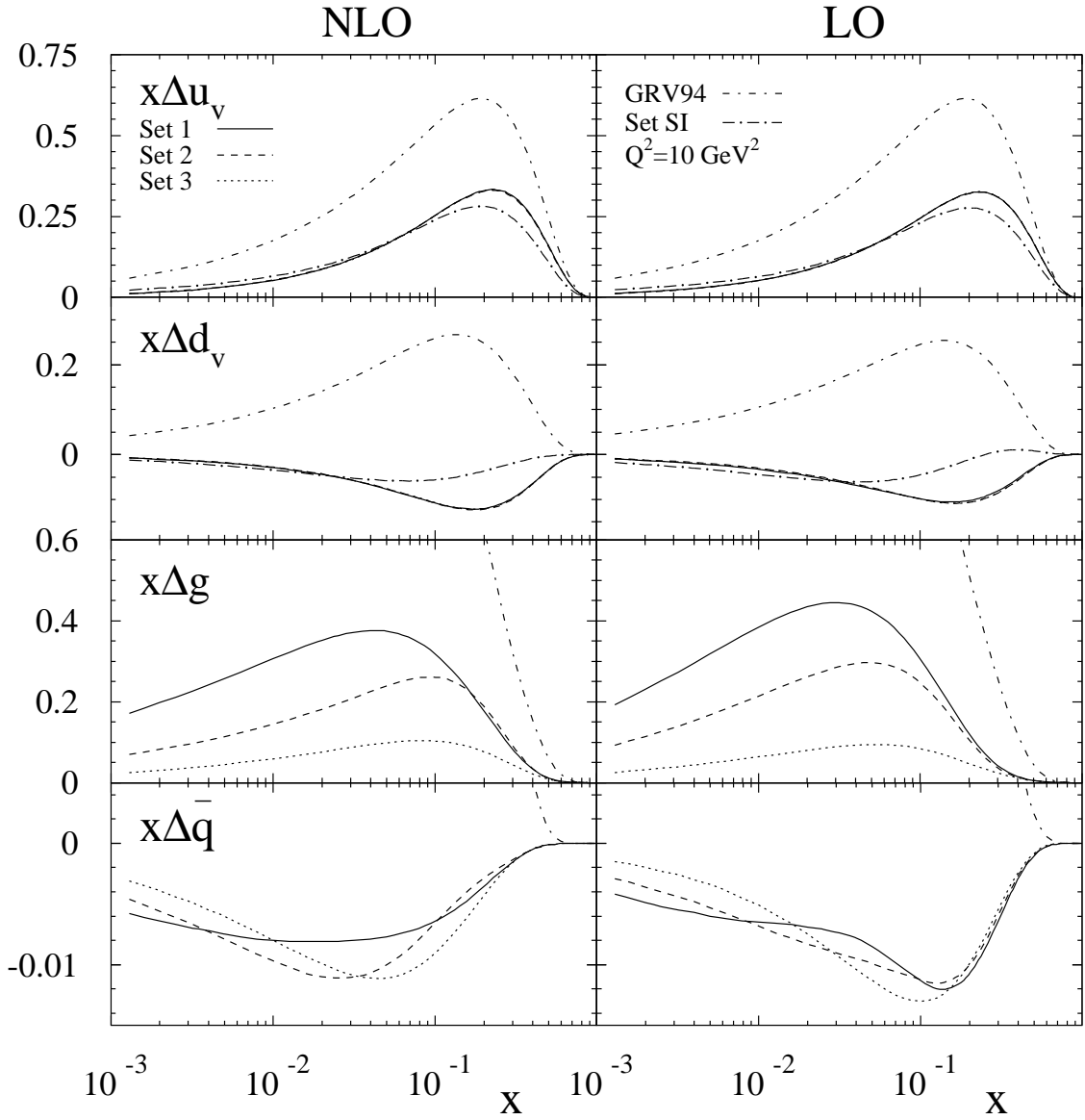,width=17.0truecm,angle=0}}
\caption{ 
Parton densities at 10 GeV$^2$.}
\end{center}
\end{figure}
\vskip 14pt
\begin{figure}[htb]
\begin{center}
\mbox{\kern0.4cm
\epsfig{file=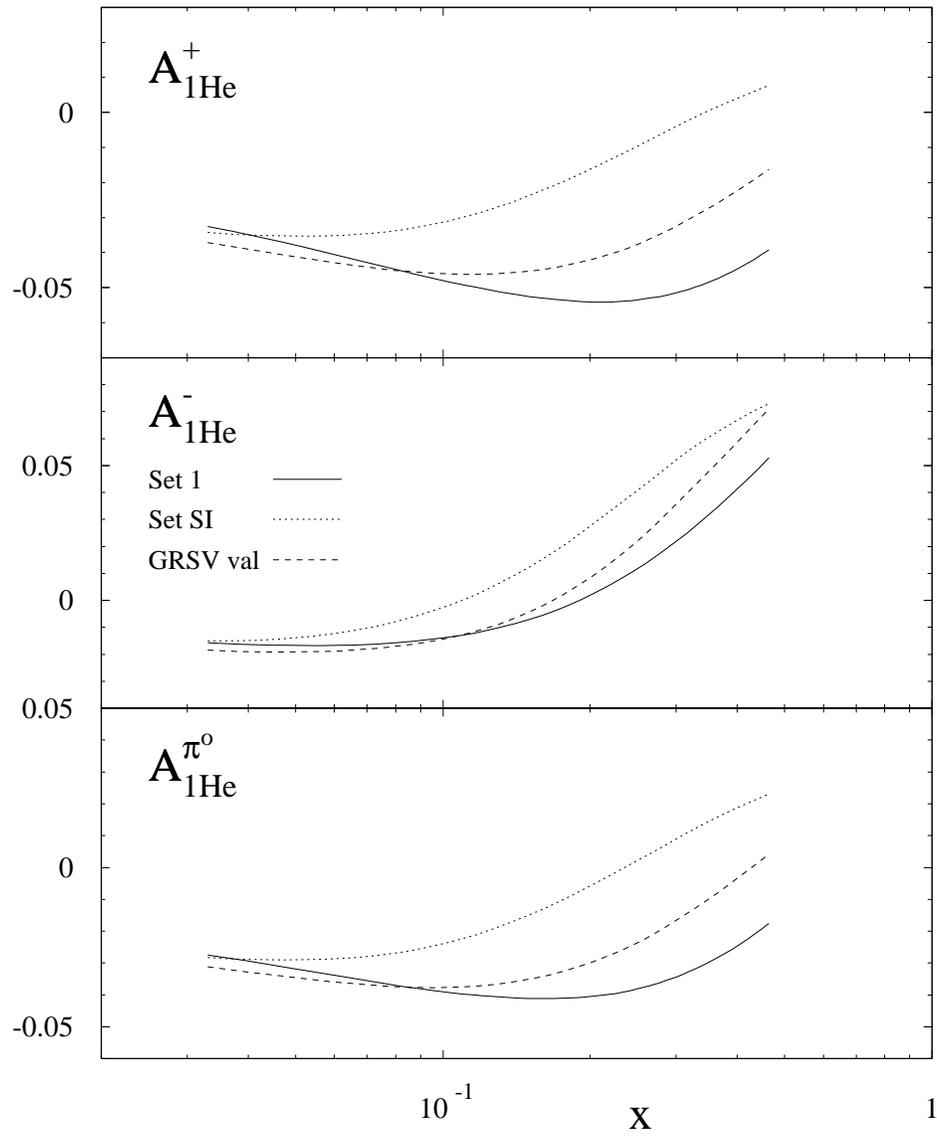,width=14.0truecm,angle=0}}
\caption{ 
Semi-inclusive asymmetries for $^3$He targets (NLO only).}
\end{center}
\end{figure}
\end{document}